\documentclass[a4paper,11pt]{article}
\pdfoutput=1 

\usepackage{jheppub} 
\usepackage{neuralnetwork}
\usepackage{caption}
\usepackage{subcaption}
\usepackage{multicol}
\usepackage{graphicx} 
\usepackage{cancel}
\usepackage{grffile}
\usepackage{braket}
\usepackage{booktabs}
\usepackage{rotating}
\usepackage{multirow}
\usepackage{algorithm}
\usepackage{algpseudocode}
\usepackage{physics} 
\usepackage{bm} 
\usepackage{url}
\usepackage[colorlinks=true,linkcolor=blue,citecolor=blue,urlcolor=blue]{hyperref}

\usepackage[normalem]{ulem}

\usepackage[T1]{fontenc} 

\preprint{ADP-23-21/T1230, JLAB-THY-23-3900}

\author[a]{N. T. Hunt-Smith,}
\author[b]{W. Melnitchouk,}
\author[b,c]{F. Ringer,}
\author[b]{N. Sato,}
\author[a]{A. W. Thomas,}
\author[a]{M.~J.~White}

\affiliation[a]{CSSM and ARC Centre of Excellence for Dark Matter Particle Physics, University of Adelaide, North Terrace, SA 5005, Australia}
\affiliation[b]{Jefferson Lab, Newport News, Virginia 23606, USA}
\affiliation[c]{Department of Physics, Old Dominion University, Norfolk, Virginia 23529, USA}
\emailAdd{nicholas.hunt-smith@adelaide.edu.au}
\emailAdd{wmelnitc@jlab.org}
\emailAdd{fmringer@jlab.org}
\emailAdd{nsato@jlab.org}
\emailAdd{anthony.thomas@adelaide.edu.au}
\emailAdd{martin.white@adelaide.edu.au}

\title{\boldmath Accelerating Markov Chain Monte Carlo sampling with diffusion models}

\abstract{
Global fits of physics models require efficient methods for exploring high-dimensional and/or multimodal posterior functions. We introduce a novel method for accelerating Markov Chain Monte Carlo (MCMC) sampling by pairing a Metropolis-Hastings algorithm with a diffusion model that can draw global samples with the aim of approximating the posterior. We briefly review diffusion models in the context of image synthesis before providing a streamlined diffusion model tailored towards low-dimensional data arrays. We then present our adapted Metropolis-Hastings algorithm which combines local proposals with global proposals taken from a diffusion model that is regularly trained on the samples produced during the MCMC run. Our approach leads to a significant reduction in the number of likelihood evaluations required to obtain an accurate representation of the Bayesian posterior across several analytic functions, as well as for a physical example based on a global analysis of parton distribution functions. Our method is extensible to other MCMC techniques, and we briefly compare our method to similar approaches based on normalizing flows. A code implementation can be found at \url{https://github.com/NickHunt-Smith/MCMC-diffusion}. }

\begin{document} 
\maketitle
\flushbottom

\section{Introduction}
\label{sec:intro}

A large number of scientific theories exist as parametric models, in which predictions depend on the values of a set of free parameters that are \emph{a priori} unknown. To progress our understanding of these theories, it is necessary to infer the values of model parameters by comparing theoretical predictions to one or more sets of observed data. Rigorous global statistical fits of theories of interest in particle astrophysics typically involve tens to hundreds of parameters (if one includes nuisance parameters), and frequently require the simulation of a wide range of data in collider physics, flavor physics, astrophysics and cosmology~\cite{DarkMachinesHighDimensionalSamplingGroup:2021wkt, Kvellestad:2019vxm}. The computational expense of this process necessitates the development of fast and accurate sampling techniques that can reduce the time required to adequately explore large parameter space volumes.

In the Bayesian statistical framework, parameter inference is performed by mapping the posterior function, defined as:
\begin{equation}
  p(\bm{\theta}|\bm{D})=\frac{\mathcal{L}(\bm{D}|\bm{\theta})\,p(\bm{\theta})}{\int \dd{\bm{\theta}}\mathcal{L}(\bm{D}|\bm{\theta})\,p(\bm{\theta})
  }\, ,
\end{equation}
where $\bm{D}$ is a set of observed data, and $\bm{\theta}$ is the set of parameters of the model. The \emph{prior} $p(\bm{\theta})$ gives the \emph{a priori} degree of belief that the parameters take certain values (often chosen to be a flat or logarithmic distribution), and $\mathcal{L}(\bm{D}|\bm{\theta})$ is the likelihood of the data given a set of parameters, which is conventionally obtained by simulation. The denominator, known as the \emph{Bayesian evidence}, is a normalisation constant that ensures that the 
function $p(\bm{\theta}|\bm{D})$ is a valid probability density function.  

Except for simple examples, the posterior is not known or easily calculable analytically. It can instead be obtained through sampling (see Ref.~\cite{https://doi.org/10.48550/arxiv.2004.06425} for a comprehensive review of algorithms). Amongst the most popular class of technique is Markov Chain Monte Carlo sampling~\cite{liu2008}. This proceeds by starting at an initial point in the parameter space and proposing new samples from a proposal density that is easy to sample, before accepting them using a condition that must satisfy strict properties. A suitable acceptance condition ensures that the samples are distributed according to the posterior density. In particle astrophysics applications, the Metropolis-Hastings algorithm is particularly popular~\cite{Metropolis1953,hastings70,MacKay2003}. 

A well-known deficiency of MCMC methods concerns multi-modal posterior functions. A proposal function based on a local proposal from the current point in the Markov Chain tends to lead to one mode being found, but often has low efficiency for jumping between modes. An uninformed, nonlocal proposal distribution, meanwhile, will typically lead to a high rejection rate. Finding more efficient proposal functions for the exploration of multi-modal posteriors reduces the time required to adequately explore the parameter space, hence allowing problems of greater complexity to be tackled.

In this paper, we present a new algorithm that augments a Metropolis-Hastings sampler with a proposal function based on a diffusion model. A diffusion model is a parametrized Markov chain trained to produce samples matching a set of input data after finite time. Such models have recently been shown to be highly effective in image synthesis, including outperforming previous approaches based on generative adversarial networks~\cite{DBLP:journals/corr/Sohl-DicksteinW15,https://doi.org/10.48550/arxiv.1907.05600,https://doi.org/10.48550/arxiv.2006.11239,https://doi.org/10.48550/arxiv.2006.09011,DBLP:journals/corr/abs-2102-09672,https://doi.org/10.48550/arxiv.2105.05233,https://doi.org/10.48550/arxiv.2112.10741,https://doi.org/10.48550/arxiv.2204.06125}. By creating a diffusion model from the available samples at regular intervals during a Metropolis-Hastings sampling run, one can develop an approximation of the posterior that can be used as an informed global proposal function to increase the efficiency of MCMC sampling. This is not the only way one could perform posterior sampling using diffusion models; one could instead use stochastic differential equations to gradually approximate the posterior via diffusion processes ~\cite{montanari2023posterior}. An alternative approach based on the use of normalizing flows for lattice QCD was previously presented in Ref.~\cite{Albergo:2019eim,Boyda:2020hsi,Albergo:2021vyo}, and a similar combination of normalizing flows with MCMC sampling was outlined in Ref.~\cite{https://doi.org/10.48550/arxiv.2107.08001,Gabrie:2021tlu}. normalizing flows have in fact been combined with diffusion models, though not in the context of MCMC~\cite{zhang2021diffusion}. We note that there are a number of important differences between diffusion models and normalizing flows:

\begin{itemize}
    \item A flow-based approach requires the estimation of gradients, while for diffusion models this is unnecessary. This is a highly useful feature when considering, for example, particle astrophysics global fits which often use stochastic simulations where gradient information is impossible to obtain~\cite{Balazs:2022tjl, GAMBIT:2023yih, Chang:2023cki}. 
    \item A fully-trained normalizing flow has direct access to a probability distribution function that approximately matches the target function~\cite{Albergo:2019eim}, while a fully-trained diffusion model can only produce samples that are approximately distributed according to the target function.
    \item Normalizing flows can be subject to topological limitations, particularly for multi-modal functions~\cite{cornish2021relaxing}. Diffusion models are more successful in matching the shape of sharply-peaked or multi-modal functions. 
    \item Diffusion models can be slow to train compared to normalizing flows given the need to train on many time-discretization steps~\cite{zhang2021diffusion}. We address this concern by fitting parameters that define the variance of the noise added in each time step, rather than training a neural network. This speeds up the training process dramatically.
\end{itemize}

We benchmark our improved technique against regular Metropolis-Hastings sampling, but one could easily accelerate other MCMC samplers using the same diffusion model technique presented here. 

This paper is structured as follows. In Section~\ref{sec:diffusion}, we provide a brief, self-contained introduction to diffusion models. In Section~\ref{sec:algorithm}, we present our algorithm that marries a diffusion model to an MCMC sampler. In Section~\ref{sec:tests_analytic}, we perform a series of numerical tests of our new algorithm by attempting to map various analytic test functions. We compare the performance of our algorithm in these tests to other basic MCMC samplers, as well as to a specific normalizing flow algorithm. In Section~\ref{sec:tests_physical}, we extend the numerical tests to a physical example relating to a global fit of parton distribution function (PDF) parameters. We summarize our conclusions in Section~\ref{sec:conclusion}.

\section{A brief introduction to diffusion models}
\label{sec:diffusion}

Diffusion models are a form of \emph{generative model} that, after training on a set of data, are able to reproduce examples of that data. Starting with, for example, an image, the key concept in a diffusion model is to systematically decay the image into pure noise through the repeated addition of a noise component (usually chosen to be Gaussian). It is then possible to learn the reverse of this noise addition process and recover the original information. Examples of valid images can then be obtained by feeding pure noise distributions through the learned reverse process.

To put this on a more precise footing, assume that we start with a data vector $\bm{x}_0$. We can then refer to a data probability distribution $q(\bm{x}_0)$, which is the distribution of possible ``real'' data vectors.  Any sample from this distribution is itself a valid data vector, and a sample from this distribution is written as $\bm{x}_0\sim q(x_0)$. We now want to add Gaussian noise for $T$ successive time steps. We can define a \emph{forward diffusion process} $q(\bm{x}_t|\bm{x}_{t-1})$ which adds the noise at time step $t$. Note that we can formally define this as a Markov chain, since the prediction of the probability density at time $t$ only depends on the immediate predecessor at time $t-1$. In principle, the width of the Gaussian noise that we add at each time step can be different, so we further define a \emph{variance schedule} $0 < \beta_1 < \beta_2 < \ldots < \beta_T <1$ where each $\beta_t$ is set to the same value across each parameter, then define the forward diffusion process as:
\begin{equation}
q(\bm{x}_t|\bm{x}_{t-1})=\mathcal{N}\big(\bm{x}_t;\sqrt{1-\beta_t}\,\bm{x}_{t-1},\beta_t\bm{I}\big).
\label{eq:addnoise}
\end{equation}
As time progresses, each new data vector at time step $t$ is drawn from a Gaussian distribution with mean $\bm{\mu}_t=\sqrt{1-\beta_t}\,\bm{x}_{t-1}$ and variance $\sigma_t^2=\beta_t$. Starting from the original vector $\bm{x}_0$, we generate a series of increasingly noisy vectors $\bm{x}_1,\ldots,\bm{x}_t,\ldots,\bm{x}_T$, where the aim is to have $\bm{x}_T$ be pure Gaussian noise.

We know the forward process very well, since it is easy to apply Gaussian noise. However, what about the reverse process? Let the probability distribution that takes us backwards along the chain of images be $p(\bm{x}_{t-1}|\bm{x}_t)$. In general, $p(\bm{x}_{t-1}|\bm{x}_t)$ cannot be known analytically, since we would need to know the distribution of all possible images in order to calculate it. Instead, we can introduce an approximation to the reverse transformation $p_\phi(\bm{x}_{t-1}|\bm{x}_t)$, with $\phi$ representing the parameters of a given model for the reverse process. The model is typically provided by a trained neural network, although in principle one could parametrize a model that appropriately describes each reverse step.

For Gaussian diffusion in the continuous limit ({\it i.e.}, in the limit of small step size, $\beta$), the reversal of the diffusion process has the identical functional form as the forward process~\cite{DBLP:journals/corr/Sohl-DicksteinW15, Feller1949OnTT}~\footnote{Note that this is also true of binomial diffusion.}. The reverse probability distribution can therefore be described by a Gaussian distribution with a mean $\bm{\mu}_\phi$ and a variance $\Sigma_\phi$:
\begin{equation}
p_\phi\big( \bm{x}_{t-1}|\bm{x}_t \big) = \mathcal{N}\big(\bm{x}_{t-1};\bm{\mu}_\phi(\bm{x}_t,t),\Sigma_\phi(\bm{x}_t,t)\big).
\label{eq:reverse}
\end{equation}
The mean and variance are also functions of the time $t$, since the reverse transformation is slightly different at each time slice. 

\section{A diffusion-model assisted Markov Chain Monte Carlo sampler}
\label{sec:algorithm}

For testing the effectiveness of diffusion model accelerated MCMC sampling, we paired the diffusion model with a simple Metropolis-Hastings (MH) MCMC algorithm. In a MH algorithm, the proposed next point $\bm{\theta}'$ in the MCMC chain is obtained by sampling from a proposal function $Q(\bm{\theta}'|\bm{\theta})$ that depends only on the previous point $\bm{\theta}$. For a given posterior~$P$, the proposed point is then accepted with probability~\cite{Allanach:2007qj}
\begin{equation}
    a = \min\left(1,\frac{P(\bm{\theta}')}{P(\bm{\theta})} \cdot \frac{Q(\bm{\theta}|\bm{\theta}')}{Q(\bm{\theta}'|\bm{\theta})}\right).
    \label{eq:MH}
\end{equation}
For $Q$ we adopt the typical choice of a Gaussian proposal function centred on $\bm{\theta}$, which as a symmetric function reduces the $Q$ term in Eq.~(\ref{eq:MH}) to unity. We identify this algorithm with a Gaussian $Q$ as \emph{pure MH}. To incorporate a diffusion model into this process, we occasionally replace the Gaussian proposal function with a diffusion model, which is periodically re-trained as the MCMC chain accumulates samples. These ``diffusion proposal samples'' are generated from random noise via the diffusion model as described in Section~\ref{sec:diffusion}, and as such each diffusion proposal sample is completely independent of the previous sample. This is a special case of MH, called \emph{independence MH}, which simplifies the $Q$ factors in Eq.~(\ref{eq:MH}) to ${Q(\bm{\theta})}/{Q(\bm{\theta}')}$~\cite{10.1214/aos/1176325750}. As long as the diffusion proposal samples are run through the MH accept/reject step, the distribution of samples is guaranteed to satisfy detailed balance and converge to the posterior eventually~\cite{mackay2003information, 10.1214/aos/1176325750}. In addition, as the overall samples get closer to the true posterior, the diffusion model is continually being trained to produce samples that more closely resemble the true posterior. Eventually, the diffusion model will be asymptotically exact --- that is, the diffusion model samples will be distributed such that they approximate the target posterior, in which case $P(\bm{\theta}) \approx Q(\bm{\theta})$~\cite{Boyda:2020hsi}. In the limit of many samples, the acceptance rate of the diffusion samples will be close to 1. One can think of this process as a method of correcting the proposal distribution to match the desired distribution $P$. 

Since the diffusion proposal samples are distributed according to our knowledge of the posterior, they will only be symmetric in the cases where the posterior is also symmetric. Therefore, in most situations the $Q$ term will not simply reduce to unity when using the diffusion model as a proposal function. We can approximate the probability of proposing a given sample $Q(\bm{\theta})$ with a Gibbs sampling method. For a given proposal vector in $D$ dimensions $\bm{\theta} = (\theta_1,\theta_2,\theta_3,...,\theta_D)$, we first draw a number of diffusion proposal samples. We then take a 1D histogram of those samples to find the approximate marginal distribution of the $\theta_1$ parameter. $Q(\theta_1)$ is then estimated by dividing the number of samples in the bin that $\theta_1$ falls into by the total number of samples. Following this step, we set $\theta_1$ at its specific value and simulate from the next parameter $\theta_2$, before taking another 1D histogram of the $\theta_2$ parameter to estimate $Q(\theta_2|\theta_1)$. We repeat this $D$ times, before multiplying the probabilities together according to Bayes' rule,
\begin{equation}
    Q(\bm{\theta}) = Q(\theta_1) \times Q(\theta_2|\theta_1) \times Q(\theta_3|\theta_1,\theta_2) \times \cdots \times Q(\theta_D|\theta_1,\theta_2,\theta_3, \ldots, \theta_{D-1}).
    \label{eq:Gibbs}
\end{equation}
This method was chosen primarily to reduce computation time and avoid having to make use of high-dimensional histograms.

Retaining a mix of diffusion model proposals and Gaussian proposals is necessary because the diffusion model can only mimic the distribution of samples it has been provided with. Hence, at early points in the MCMC run, the diffusion model will not provide an adequate approximation of the final posterior, and therefore samples taken from it will not have a particularly high acceptance rate. Later in the run, however, the diffusion model will provide a more accurate representation of the posterior, and the acceptance rate will improve. Inspired by the algorithmic structure in Refs.~\cite{https://doi.org/10.48550/arxiv.2107.08001, Gabrie:2021tlu}, we therefore designed Algorithm~\ref{alg:1}, which uses pure MH some of the time alongside the diffusion proposal samples.

\begin{algorithm}
\caption{}\label{alg:1}
\begin{algorithmic}[1]
\For{$i = 1,...,n$}
    \State Generate a random number $r$ from the uniform distribution $\mathcal{U}(0,1)$
    \If{$r < p_{\rm diff}$}
        \State{Metropolis-Hastings with diffusion model proposal}
    \Else
        \State{Pure Metropolis-Hastings with Gaussian proposal}
    \EndIf

    \If{$i \mod \tau = 0$}
        \State{Retrain diffusion model on existing samples}
    \EndIf
\EndFor
\end{algorithmic}
\end{algorithm}

In Algorithm~\ref{alg:1}, $n$ is the total number of samples to be generated, $p_{\rm diff}$ is the probability of using a diffusion proposal sample as opposed to a local Gaussian proposal sample, and $\tau$ is the number of samples to generate before retraining the diffusion model. $\tau$ is tunable to the posterior one wishes to sample from, with lower values of $\tau$ resulting in better performance at the cost of computation time to retrain the diffusion model. 

The primary advantage of using the diffusion model over standard MCMC is the nonlocal nature of its proposed points within a single chain. Where a standard MCMC chain might get stuck in a local mode, the diffusion model can jump to any point in parameter space. Algorithms such as Explore-Exploit MCMC (Ex$^2$MCMC)~\cite{NEURIPS2022_21c86d5b} and normalizing flows (described in Sec.~\ref{sec:intro}) make use of this global transition kernel. For multi-modal posteriors, this means one can accurately characterize the relative weights between the modes with only a single chain. Our only requirement occurs if the modes are distanced too far apart for the Gaussian proposal to jump between them with a significant probability. In that case, all the important modes must be located beforehand in order to seed the diffusion model with points at those modes for initial training, allowing transitions between the modes. 

The diffusion model can also accelerate the MCMC process even for posteriors with a single mode. The proposed diffusion samples are more likely to be located in strongly peaked regions of the parameter space, meaning that if the MCMC chain strays far away from the mode the diffusion samples will bring it back to the region of interest. This is particularly significant for computationally expensive posteriors, where reducing the number of likelihood evaluations is the highest priority. A further advantage of the diffusion model is that, once the sampling algorithm has converged, the diffusion model has already been trained to produce samples that closely resemble the posterior. We therefore obtain a fast way of generating as many approximate posterior samples as desired. This is particularly useful in applications where a rough initial scan can be used to define a much smarter prior for a subsequent scan. Global fits of parameter spaces which exhibit a degree of fine-tuning are an obvious use-case. 

Applying a diffusion model to MCMC samples rather than images is a considerably simpler task in terms of algorithmic complexity. A single high-quality image consists of millions of pixels in several color channels which must be converted into a high-dimensional tensor, and millions of training images must be supplied in order for the diffusion model to learn and then generate an image that looks somewhat like the training images. On the other hand, our MCMC samples are simple low-dimensional arrays with a length equal to the number of dimensions of the problem; rarely more than a few hundred for realistic physics examples. We have therefore constructed a diffusion model that is tailored towards MCMC samples rather than attempting to adapt existing image-based diffusion models.

When we are training the diffusion model, we define the forward Gaussian noise process by adding noise to the samples in the same way as Eq.~(\ref{eq:addnoise}). In other words, each vector in the overall set of samples that form the Markov chain is corrupted slightly at each time step. This forward process will be concluded once the distribution of the vectors has been converted from a shape that resembles the posterior to pure Gaussian noise. This introduces two hyperparameters which can be tuned as appropriate: the number of time steps and the amount of noise progressively added at each time step, $\beta_T$. Generally, increasing these values will further reduce the minimum loss, but also increase the computation time needed to find the minimum.

To train the diffusion model, we first perform the forward diffusion process according to Eq.~(\ref{eq:addnoise}) until we are left with pure Gaussian noise. On an individual sample level,
\begin{equation}
\bm{x}_t = \sqrt{1-\beta_t}  \bm{x}_{t-1} + \sqrt{\beta_t} \bm{\epsilon},
\label{eq:forward}
\end{equation}
where $\bm{\epsilon} \sim \mathcal{N}(0,\bm{I})$. We then define a new data vector $\bm{y}_T$ drawn from pure Gaussian noise with the same length as $\bm{x}_T$. Each reverse step is then computed using
\begin{equation}
\bm{y}_{t-1} = \bm{y}_{t} - (\bm{x}_{t} - \bm{x}_{t-1}) \bm{\phi}_{t}.
\label{eq:train}
\end{equation}
Here, the difference in samples between each time step $(\bm{x}_{t-1} - \bm{x}_{t})$ is multiplied by a free parameter $\bm{\phi}_t$ to be trained. Repeating this operation $T$ times leads to the denoised samples $\bm{y}_0$. We can define a loss function $L$ as
\begin{equation}
L = \sum_{t} \big(\bm{y}_{t-1}(\bm{\phi}_{t}) - \bm{x}_{t-1}\big)^2,
\end{equation}
summed over the total number of samples. Rather than training a neural network, we therefore perform a series of fits of the $\bm{\phi}_t$ parameters to minimise the sum of squared differences in each time step between the denoised samples we have generated and the original MCMC samples. There are $T \times D$ parameters to be fit simultaneously for each retrain of the diffusion model. To minimise the loss function we use the \verb|SciPy minimize| function.

We then perform the reverse process by generating another new data vector $\bm{z}_T$ from pure Gaussian noise, before applying Eq.~(\ref{eq:train}) to obtain $\bm{z}_0$. We can repeat this process indefinitely, generating as many samples as required almost instantaneously. This reverse process is considerably simpler than the typical method of parametrizing reverse Gaussian transformations described in Eq.~(\ref{eq:reverse}). Empirically, we found it to be highly effective at generating unique diffusion samples approximately distributed according to the target distribution.

\section{Analytic test functions}
\label{sec:tests_analytic}

In the test functions below, we explore the advantages Algorithm~\ref{alg:1} has over pure MH, as well as over a normalizing flow MCMC algorithm. The values assigned for each hyperparameter in each function are contained in Appendix~\ref{app:params}. A further test comparing Algorithm~\ref{alg:1} and pure MH for the 4D Rosenbrock function can be found in Appendix~\ref{app:Rosenbrock}.

\subsection{2D Himmelblau function}

\begin{figure}[t]
\centering
\includegraphics[scale = 0.5]{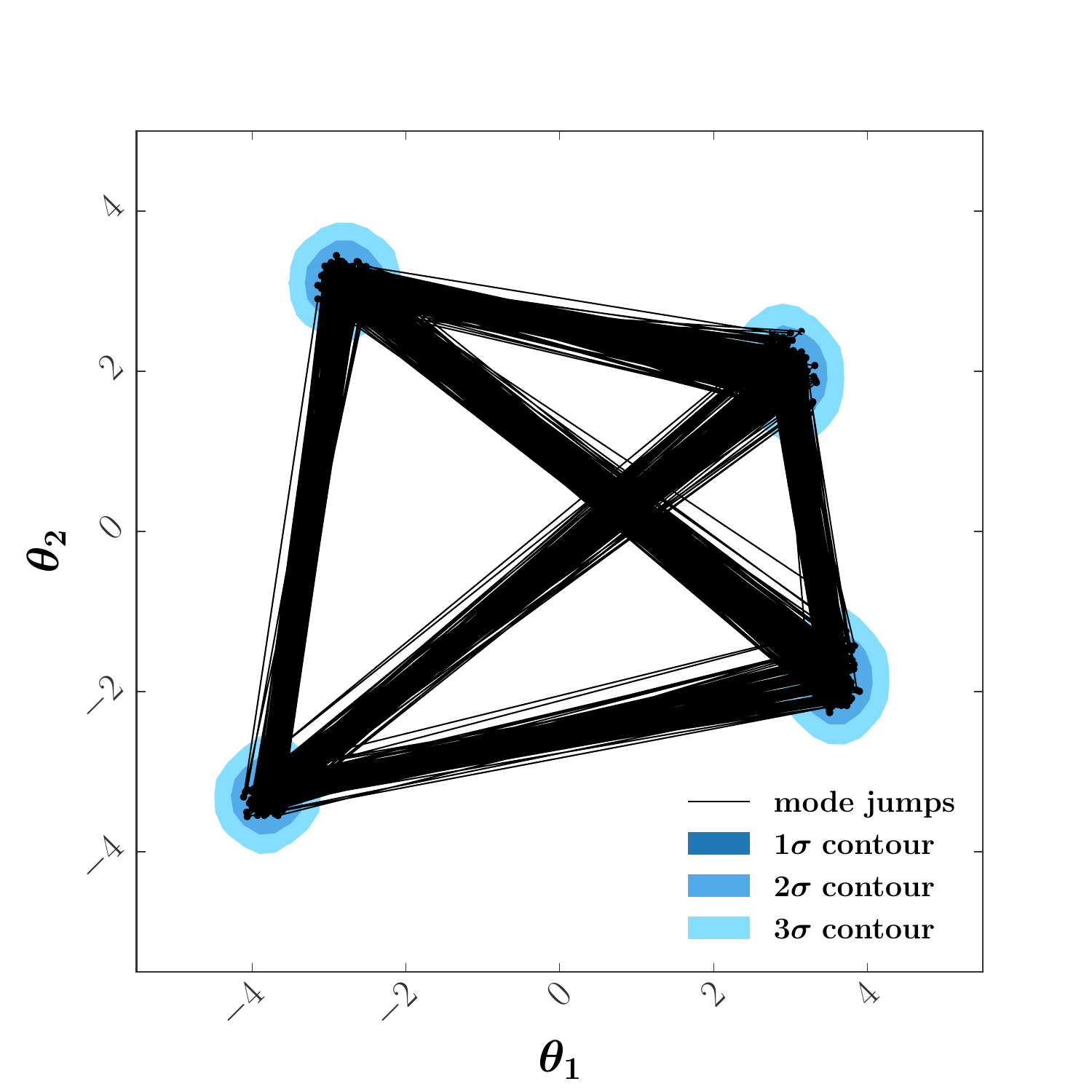}
\caption{2D Himmelblau function sampled with Algorithm~\ref{alg:1} (10,000 samples). Black lines depict jumps between modes due to the diffusion model.}
\label{fig:Himmelblau}
\end{figure}

The Himmelblau function is given by
\begin{equation}
    f(\bm{\theta}) = (\theta_1^2 + \theta_2 - 11)^2 + (\theta_1 + \theta_2^2 - 7)^2,
\end{equation}
which consists of 4 sharply peaked modes of equal height spaced a large distance away from each other, presenting a significant challenge for MCMC samplers. Without multiple chains, we have found that a simple MCMC algorithm like pure MH will either get stuck in one of the modes if the Gaussian proposal is too narrow, or have a very low acceptance rate if the Gaussian proposal is too wide. Augmenting the MH sampler with a diffusion model as in Algorithm~\ref{alg:1} allows the chain to jump between these modes, while accurately characterizing the relative weights between the modes. This behavior is shown in Fig.~\ref{fig:Himmelblau}, which depicts 10,000 samples of the 2D Himmelblau from a single chain using Algorithm~\ref{alg:1}. The nonlocal jumps made by the diffusion model are also represented visually by the black lines connecting the 4 modes, of which there are around 1,000 in this case. The jumps between modes ensure that the posterior is very well described, with all of the modes mapped out equally while requiring relatively few function evaluations. 

Before the chain began, the diffusion model was seeded by providing 50 points located near each of the modes. Given that the modes are positioned far from each other, it is necessary to do this to allow the diffusion model to propose jumps between modes at the start. These initial points are increasingly unlikely to be proposed as the chain progresses since there are many more samples being generated, and they are removed at the end of the chain.

\subsection{10D Gaussian mixture}

\begin{figure}[t]
\centering
\includegraphics[scale = 1.2]{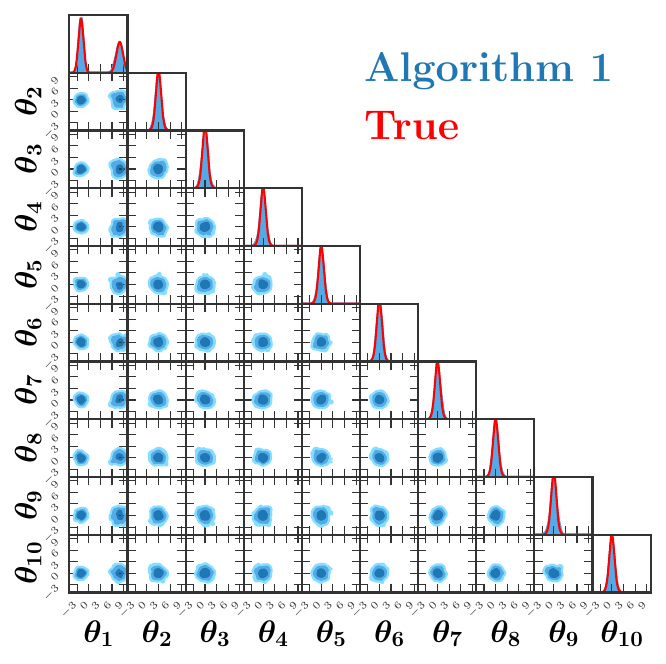}
\caption{Corner plot of Algorithm~\ref{alg:1} for a 10D Gaussian mixture with 2 components (5,000 samples). The true marginalized posteriors are shown in red, blue contours represent sample density.}
\label{fig:Gaussian}
\end{figure}

Here we compare our results using Algorithm~\ref{alg:1} to results obtained from a normalizing flow MCMC algorithm for a high-dimensional Gaussian mixture function with 2 components. In the same way as in~\cite{https://doi.org/10.48550/arxiv.2107.08001}, we define a mixture of 2 Gaussians in 10D,
\begin{equation}
    f(\theta_i) = w_A \frac{e^{-\frac{1}{2}|\theta_i - \theta_{A,i}|^2}}{(2 \pi)^{d/2}} + w_B \frac{e^{-\frac{1}{2}|\theta_i - \theta_{B,i}|^2}}{(2 \pi)^{d/2}},
\end{equation}
with weights $w_A = 2/3$, $w_B = 1/3$ and non-zero means only in the first two dimensions $\theta_{A 1,2} = (8,3)$, $\theta_{B 1,2} = (-2,3)$. We therefore have to sample two modes in the first two dimensions situated significantly far away from one another, with different relative weights. As shown in Fig.~\ref{fig:Gaussian}, Algorithm~\ref{alg:1} converges to the truth within 5,000 samples, which is very few function evaluations given the high-dimensional nature of this problem. Once again, the diffusion model was initially seeded with 50 points near the two modes. 

\begin{figure}[t]
\centering
\includegraphics[scale = 0.75]{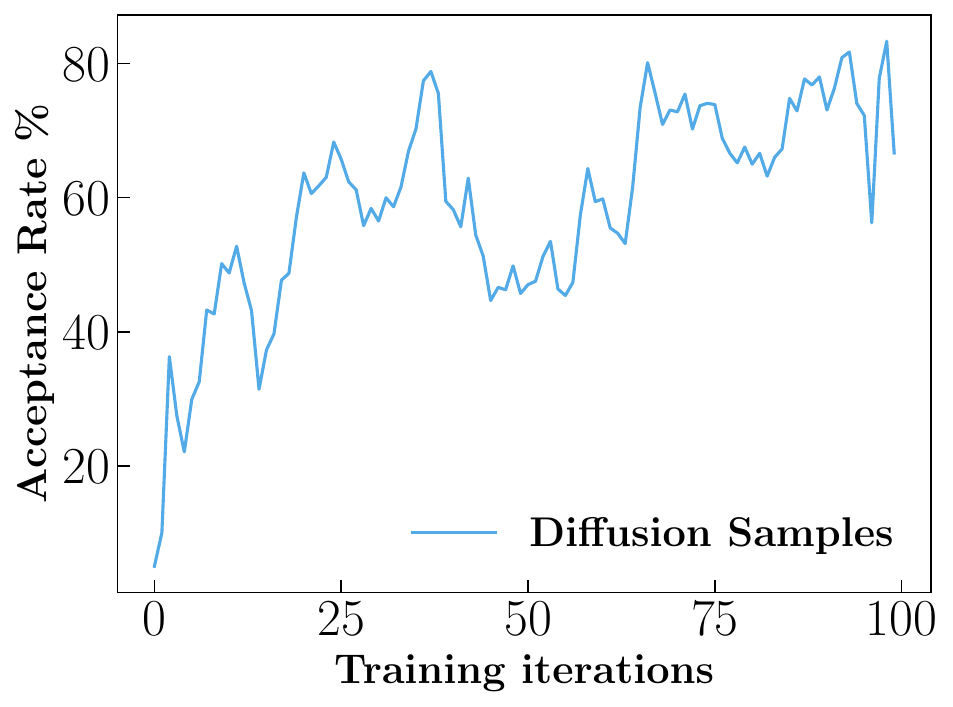}
\caption{ Acceptance rate with Algorithm~\ref{alg:1} for a 10D Gaussian mixture with 2 components.}
\label{fig:diff_vs_norm}
\end{figure}

Importantly, our diffusion model MCMC sampler outperforms a similar normalizing flow MCMC sampler for this same problem by requiring fewer training iterations in order to reach a high acceptance rate. Figure~\ref{fig:diff_vs_norm} shows that Algorithm~\ref{alg:1} is able to reach close to 80\% acceptance within less than 100 training iterations, whereas the flow-based approach requires at least 500 iterations to reach a comparable acceptance rate~\cite{https://doi.org/10.48550/arxiv.2107.08001}. 

We do however point out that neither the total number of samples nor the number of samples for each training iteration were specified for the normalizing flow algorithm. We further note that the full structure of the normalizing flow algorithm is significantly more complex than our Algorithm~\ref{alg:1}. A normalizing flow is an invertible map that takes a base density such as a Gaussian with unit variance, and pushes it forward towards a target density. In~\cite{https://doi.org/10.48550/arxiv.2107.08001}, neural networks are trained to learn the mapping between the base and target densities by minimizing the Kullback-Leibler divergence. The normalizing flow is therefore supervised at the PDF level rather than the sample level, which may explain the increased performance on the side of the diffusion model. 

Where we have paired our diffusion model with a basic local Gaussian transition kernel, the normalizing flow algorithm is combined with a more sophisticated local MALA sampler. Furthermore, the diffusion model algorithm operates in a single chain, whereas hundreds of independent walkers are initialized and split between each mode at the start of the normalizing flow algorithm. We therefore expect our results to be conservative in performance - a more sophisticated transition kernel and algorithmic structure would likely result in better performance in our case. The acceptance rate comparison is merely suggestive of an advantage the diffusion model may have over normalizing flows for this particular example, an exact comparison is considered beyond the scope of this paper.

Figure~\ref{fig:diff_vs_norm} also shows the property of asymptotic exactness shared by independent MH samplers such as these. As the MCMC chain progresses and the diffusion model/normalizing flow is repeatedly retrained, the acceptance rate increases well beyond 70\%. This indicates that the proposal diffusion model is closely approximating the target posterior, as explained in Sec.~\ref{sec:algorithm}. It also highlights an important feature of algorithms of this kind - at the conclusion of the MCMC chain, one obtains a very fast approximation to the posterior from which one can draw as many samples as required. 

\subsection{4D EggBox function}

\begin{figure}[t]
\centering
\includegraphics[scale = 1]{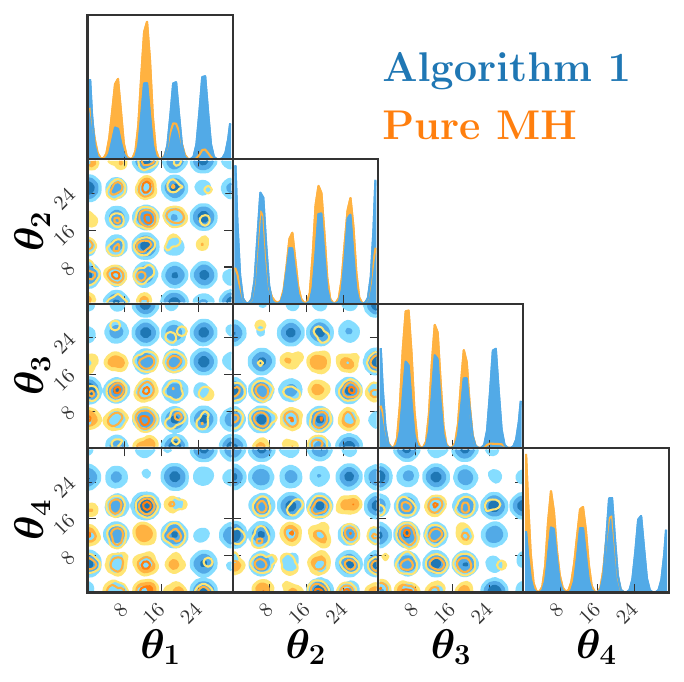}
\caption{Comparison of Algorithm~\ref{alg:1} and pure MH for the 4D EggBox function (0.1M samples). Coloured contours represent sample density for each algorithm, with more identified modes representing greater success at characterising the shape of the true posterior.}
\label{fig:EggBox}
\end{figure}

The EggBox function is given by
\begin{equation}
    f(\bm{\theta}) = \left(2 + \prod_{i = 1}^D \left[\cos\left(\frac{\theta_i}{2}\right)\right]\right)^{5},
\end{equation}
which consists of many local modes of equal height spread at regular intervals throughout parameter space. The challenge for an MCMC algorithm is to find all of the modes while weighting them equally. Given a wide enough Gaussian proposal, pure MH is eventually able to successfully find all the minima, however the inclusion of the diffusion model accelerates this process dramatically. Figure~\ref{fig:EggBox} shows a comparison between pure MH and Algorithm~\ref{alg:1} for the heavily multi-modal 4D EggBox function over 0.1M samples. For such a low number of samples, pure MH is unable to map a substantial fraction of the total modes (it produced samples in around 55\% of modes), particularly at the edges of the parameter space. On the other hand, Algorithm~\ref{alg:1} is able to find most of the modes (samples produced in around 80\% of modes), with the 1D marginal distributions showing roughly equal peaks across each parameter compared to pure MH. We found that pure MH would have to perform at least 5 times as many function evaluations to obtain similar samples as for Algorithm~\ref{alg:1} in Fig.~\ref{fig:EggBox}, demonstrating a significant acceleration of the MCMC sampling process in this case. In this case, the diffusion model was seeded by providing 1,000 random points drawn from a uniform distribution.

\begin{figure}[t]
\centering
\includegraphics[scale = 1]{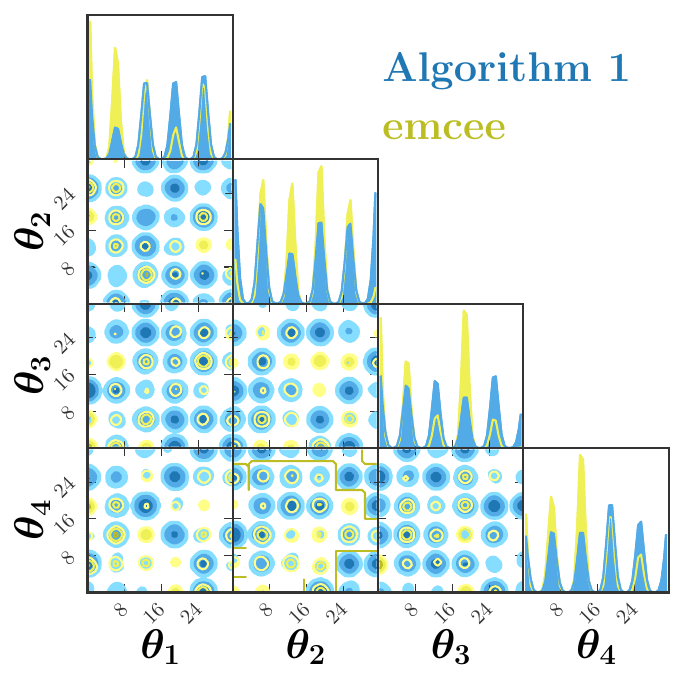}
\caption{Comparison of Algorithm~\ref{alg:1} and the emcee algorithm for the 4D EggBox function (0.1M samples). Coloured contours represent sample density for each algorithm, with more identified modes representing greater success at characterising the shape of the true posterior.}
\label{fig:EggBox_emcee}
\end{figure}

It is also worth comparing our diffusion-accelerated MCMC algorithm to an algorithm that makes use of multiple chains, such as the widely-used \verb|emcee| algorithm~\cite{Foreman_Mackey_2013}. After running the default \verb|emcee| algorithm on the EggBox function with 8 chains for a total of 0.1M samples, we created a comparison plot with Algorithm~\ref{alg:1}, see Fig.~\ref{fig:EggBox_emcee}. We found that there was only a slight improvement over the single chain of pure MH, with around 60\% of modes found. This demonstrates that Algorithm~\ref{alg:1} can outperform algorithms with multiple chains for heavily multi-modal functions.

\section{Physics example: global parton distribution function fit}
\label{sec:tests_physical}

We now apply Algorithm~\ref{alg:1} and pure MH to a simple physics example intended to reflect the salient features of a global PDF fit. PDFs describe the probability of finding a particular parton (quark or gluon) in a hadron as a function of the energy-momentum fraction $x$ carried by the parton and the energy scale $Q$ for a given collision. By parametrizing the PDFs, we can make theoretical predictions for physical observables (typically cross sections) and compare to experimental data. A global PDF fit then involves constructing a goodness of fit measure, such as the $\chi^2$, that describes the difference between theory and experiment, from which the likelihood can be defined as $\mathcal{L} = \exp(-\frac12\chi^2)$. This likelihood is highly computationally expensive to evaluate for real data, and global PDF analyses often use frequentist statistics to construct approximate uncertainty bands on the PDFs through repeated maximization of the likelihood. However, in principle a purely Bayesian approach, such as a Metropolis-Hastings sampler, would also provide valid uncertainty bands for this likelihood given enough time. For our simple PDF example, we will use a likelihood function that is much faster to evaluate in order to compare the uncertainty bands obtained via Algorithm~\ref{alg:1} and pure MH. 

The details of the construction of this simple PDF example can be found in Ref.~\cite{Hunt-Smith:2022ugn}; here, we summarize the most important elements. We first construct a set of two toy ``quark'' PDFs with $x$ dependence parametrized by a basic functional form,
\begin{equation}
    \begin{aligned}
     q_i(x) = N_i\, x^{\alpha_i} (1-x)^{\beta_i},\qquad i=1,2.
    \end{aligned}
    \label{eq:toymodel}
\end{equation}
These functions could, for example, represent the $u$ and $d$ quark PDFs in the proton, with free parameters $\alpha_i$ and $\beta_i$. We can then model toy ``cross sections'' $\sigma_j$ using a linear combination of the two PDFs,
\begin{equation}
    \begin{aligned}
    \sigma_j = \sum_{i=1,2} c_{ji}\, q_i.
    \end{aligned}
    \label{eq:toysigma}
\end{equation}
These could, for instance, correspond to inclusive deep-inelastic scattering (DIS) cross sections for a proton and a neutron, with the coefficients $c_{ij}$ proportional to the squares of the quark charges, in which case we would assign
    $c_{11} = 4 c_{12} = 4 c_{21} = c_{22}$.
The values of $x$ are taken in the range $x=0.1-0.9$ at regular intervals, depending on the number of toy data points we wish to generate. 
For each $x$ value we calculate the corresponding $q_1$ and $q_2$ PDFs, from which the true $\sigma_1$ and $\sigma_2$ are determined. 
Toy cross section data points are then generated by drawing randomly from a Gaussian distribution centred on the true cross section values, and uncertainties are assigned to be 0.1 times the magnitude of each toy data point. 

To fit this pseudodata, we define a 4D model following the traditional $\sim x^{\alpha} (1-x)^{\beta}$ parametrization, which matches the true underlying law in Eq.~(\ref{eq:toymodel}). We can then define 
\begin{equation}
    \chi^2 = \sum_{i = 1,2}\sum_j^{n_{\rm data}}
    \left( \frac{\sigma^{\rm data}_i(x_j) - \sigma^{\rm model}_i(x_j,\bm{\theta})}
                {\Delta\sigma_i(x_j)}
    \right)^2,
    \label{eq:chi2min}
\end{equation}
where $i$ is the index of the two cross sections, $j$ is the index of the data points summing to a total number of points $n_{\rm data}$ in the sample, $\sigma^{\rm data}_i$ is the toy cross section data with uncertainty $\Delta\sigma_i$, and $\sigma^{\rm model}_i$ is the model cross section [Eq.~(\ref{eq:toysigma})] for a given set of parameters~$\bm{\theta}$. The likelihood is then given by
\begin{equation}
    \mathcal{L} = \exp(-\frac12 \chi^2).
\end{equation}
Using this likelihood, we can now sample the posterior using Algorithm~\ref{alg:1} and pure MH, and generate uncertainty bands for the cross sections and PDFs. The likelihood in this case has a single mode.

\begin{figure}[t]
\centering
\includegraphics[scale = 0.5]{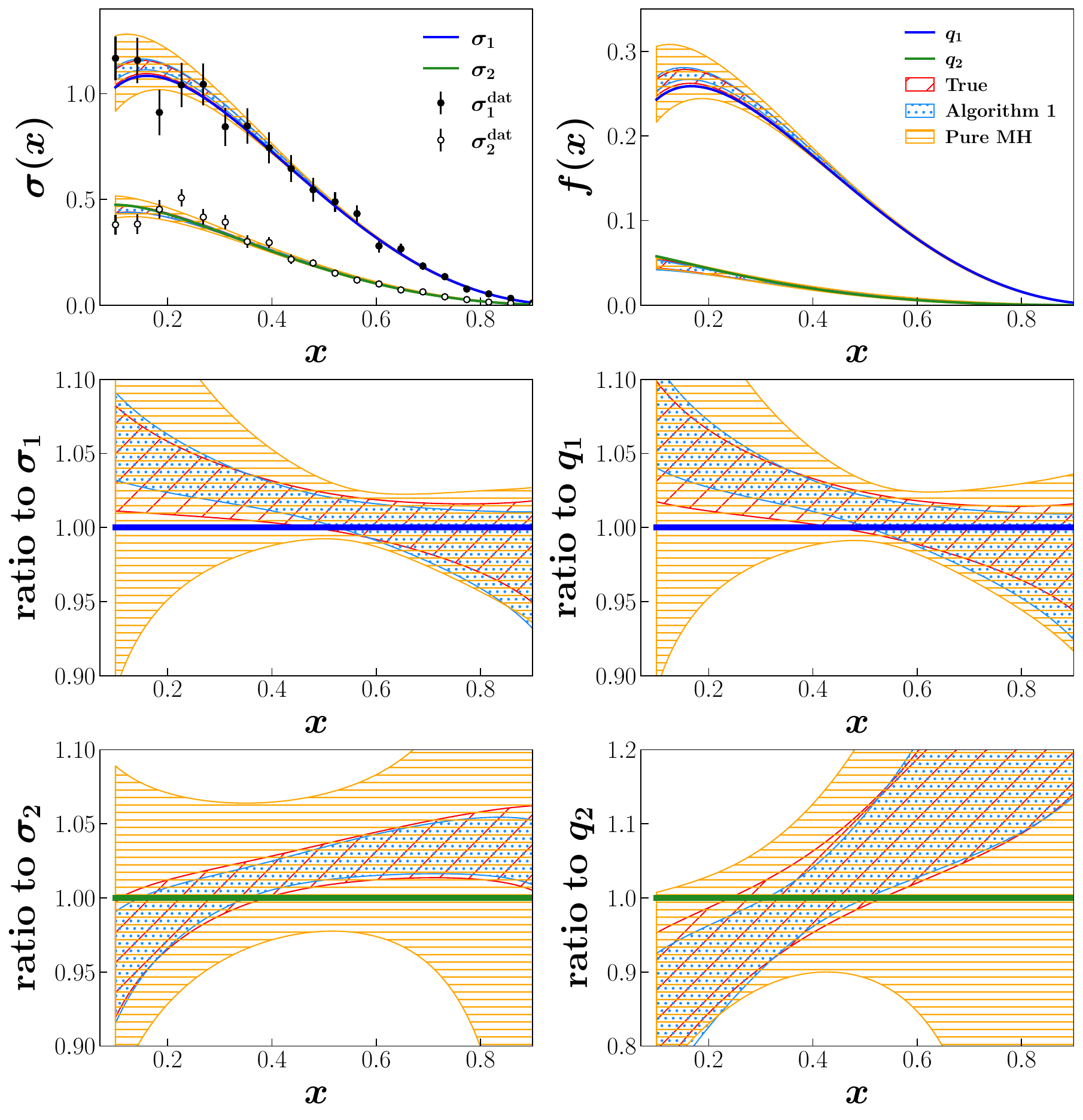}
\caption{Comparison of fit results for Algorithm~\ref{alg:1} and pure MH (10,000 samples) to the many-sample limit (1M samples). The top row contains cross section and parton distributions, while the middle and bottom rows show ratios to the true values for each distribution.}
\label{fig:PDFs}
\end{figure}

Figure~\ref{fig:PDFs} shows a comparison between Algorithm~\ref{alg:1} and pure MH for our 4D PDF physics example. We also include the uncertainty bands generated in the limit of many samples, as an estimate of the true uncertainties. We found that both Algorithm~\ref{alg:1} and pure MH agree in this limit, meaning that our diffusion-augmented algorithm does converge to the purely Markovian distribution for this example. In general, one ought to be careful when using past samples taken from the existing chain as inputs to generate new samples as we do in Algorithm~\ref{alg:1}. For example, adaptive MCMC algorithms require certain conditions to be met in order to ensure ergodicity of the chain, see~\cite{4d9c5a572a6145e9bd69003f715c3dfc} for further details.

For the 10,000 samples shown here, pure MH heavily overestimates the uncertainty bands. On the other hand, pairing MH with the diffusion model results in samples that approximate the true uncertainty bands very closely. 3 times as many samples would be required for pure MH to produce a fit of comparable quality. One advantage the diffusion model has over pure MH is that it can make nonlocal jumps to key regions of interest and avoid getting stuck in the tails of the distribution, which explains why Algorithm~\ref{alg:1} is able to converge to the truth in fewer likelihood evaluations even for a function with a single mode. The diffusion model was initially seeded with 100 points located near the mode.

\section{Conclusions}
\label{sec:conclusion}

We have developed a diffusion model tailored for low-dimensional data, pairing it with a Metropolis-Hastings sampler to create a new accelerated Markov Chain Monte Carlo algorithm. We find that Algorithm~\ref{alg:1} is many times faster than pure Metropolis-Hastings at obtaining accurate posterior samples for test functions with any number of modes, as evidenced by a significant reduction of the number of likelihood evaluations required to adequately explore the Himmelblau, 10D Gaussian mixture, and EggBox functions. Algorithm~\ref{alg:1} has also shown significant improvement over a normalizing flow algorithm for a specific test function. In the context of a real physical example based on a global analysis of parton distribution functions, Algorithm~\ref{alg:1} is able to characterize uncertainty bands much more accurately than pure Metropolis-Hastings for a limited number of samples. 

Our technique does not require gradient information, and is therefore applicable to a wide range of posteriors encountered in physics applications. Furthermore, our proof of principle indicates that a combination of a diffusion model with a more sophisticated MCMC or other sampler could be expected to yield even better results.

\newpage
\section*{Acknowledgements}
This research was supported by the ARC Centre of Excellence CE200100008 and by the DOE with contract No. DE-AC05-06OR23177, under which Jefferson Science Associates, LLC operates Jefferson Lab. The work of N.S. was supported by the DOE, Office of Science, Office of Nuclear Physics in the Early Career Program. We wish to thank Andrew Fowlie for multiple useful insights regarding Gibbs sampling and MCMC algorithms.

\appendix
\section{Hyperparameters}
\label{app:params}

Here we outline each of the hyperparameters in Algorithm~\ref{alg:1}. If their values varied between test problems, the different settings are contained in Table~\ref{table:params}.

\begin{itemize}
    \item The total number of retrains was set to 100 for each test problem.
    \item $\tau$ is the number of samples per retrain of the diffusion model. $\tau$ multiplied by (retrains) gives $n$, the total number of samples generated.
    \item $p_{\rm diff}$ is the probability of drawing a diffusion proposal sample rather than a Gaussian proposal sample. 
    \item $\sigma_{MH}$ is the width of the Gaussian proposal distribution.
    \item $\beta$ is the amount of noise progressively added in the forward diffusion process, and is set to increase linearly from 0.1 to 0.3 for each test problem.
    \item \verb|nsteps| is the number of noising steps in the forward/reverse diffusion process, and is set to 20 for each of the test problems.
    \item \verb|bins| is the number of bins in the 1D histograms when calculating Eq.~(\ref{eq:Gibbs}), and is set to 20 for each test problem. \\
\end{itemize}

\begin{table}[h]
\caption{Algorithm~\ref{alg:1} hyperparameter values assigned for each test problem.}
\begin{center}
\begin{tabular}{c|c|c|c|c}
    ~Hyperparameter & ~2D Himmelblau & ~10D Gaussian & ~4D EggBox & ~4D PDF\\
    \hline 
    ~$\tau$~ & 100 & 500 & 1000 & 100~~\\
    ~$p_{\rm diff}$~ & 0.83 & 0.1 & 0.5 & 0.5~~\\
    ~$\sigma_{\rm MH}$~ & 0.15 & 0.5 & 0.6 & 0.1~~\\
\end{tabular}
\label{table:params}
\end{center}
\end{table}

\begin{figure}[t]
\centering
\includegraphics[scale = 0.75]{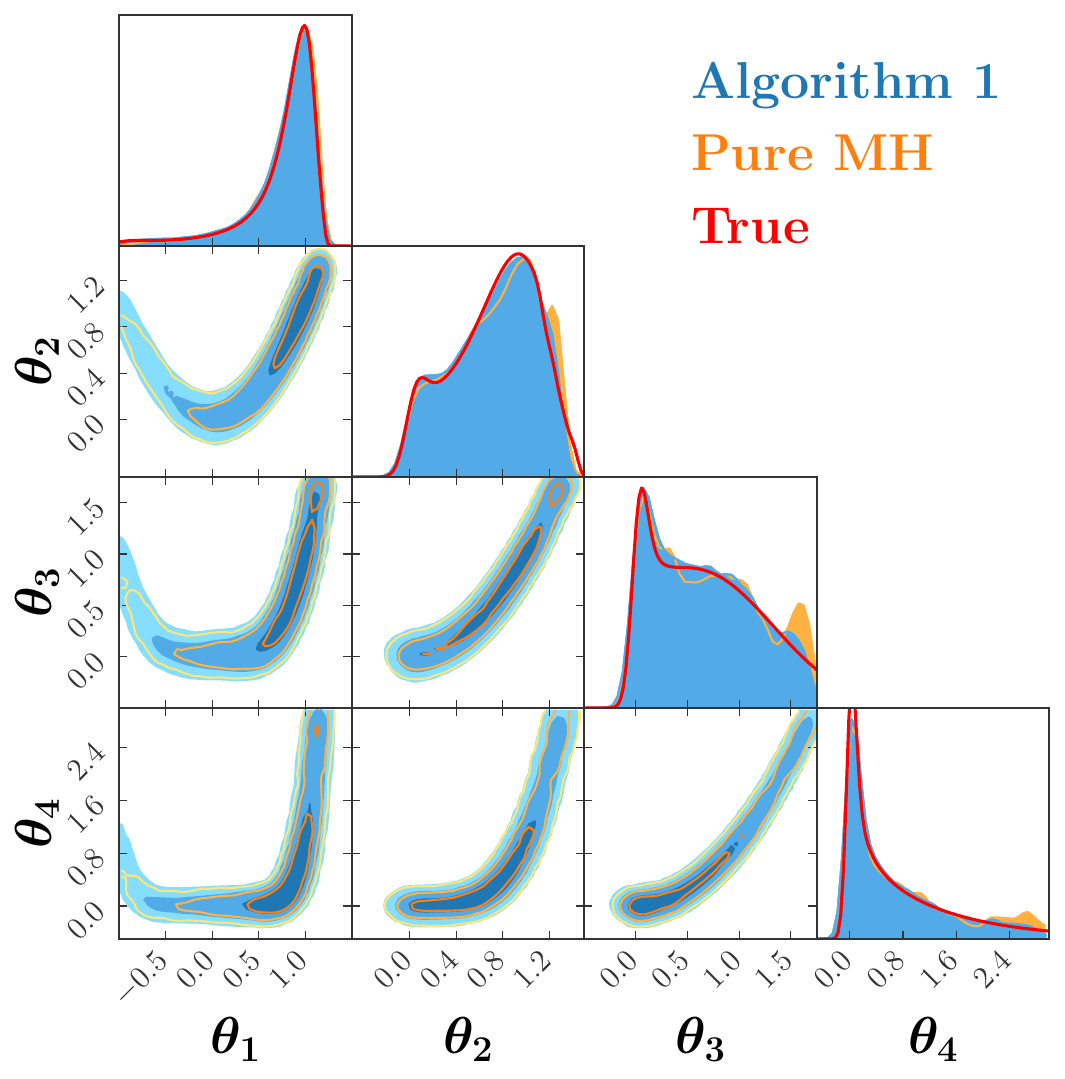}
\caption{Comparison of Algorithm~\ref{alg:1} and pure MH for the 4D Rosenbrock function (0.5M samples). The true marginalized posteriors are shown in red, coloured contours represent sample density.}
\label{fig:Rosenbrock}
\end{figure}

\newpage
\section{4D Rosenbrock function}
\label{app:Rosenbrock}

The Rosenbrock function is given by
\begin{equation}
    f(\bm{\theta}) = \sum_{i=1}^{D-1}[100 \times (\theta_{i + 1} - \theta_i)^2 + (1-\theta_i)^2],
\end{equation}
where $D$ is the number of dimensions. This banana-shaped function consists of a single mode located within a long, flat bump. In 4D, it is particularly difficult to find this mode while also correctly mapping the curved shape of the surrounding area. Figure~\ref{fig:Rosenbrock} shows a comparison between pure MH and Algorithm~\ref{alg:1} for the 4D Rosenbrock function over 0.5M MCMC samples. Pure MH produces a false  secondary mode, which is not present once the diffusion model is added in. Twice as many samples were required for pure MH to match Algorithm~\ref{alg:1} for this test function. One advantage the diffusion model has over pure MH is that it can make nonlocal jumps to key regions of interest and avoid getting stuck in the tails of the distribution, which explains why Algorithm~\ref{alg:1} is able to converge to the truth in fewer function evaluations. 

\begin{figure}[t]
\centering
\includegraphics[scale = 0.75]{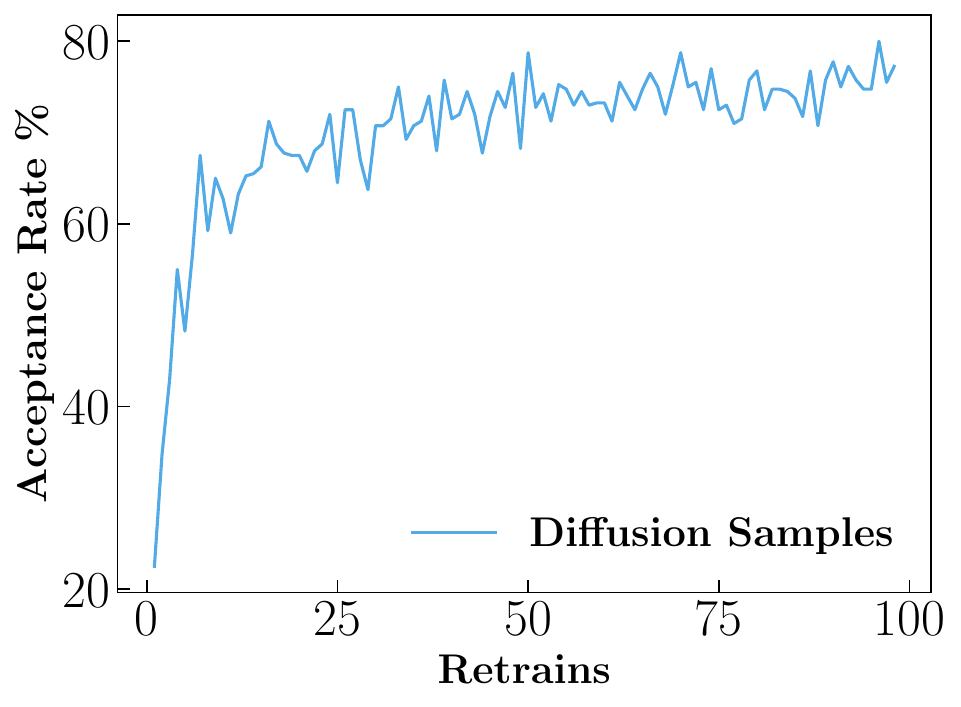}
\caption{Increase in diffusion sample acceptance rate with number of retrains for the 4D Rosenbrock function (0.5M samples).}
\label{fig:Acceptance}
\end{figure}

We also show in Fig.~\ref{fig:Acceptance} a plot demonstrating asymptotic exactness for the Rosenbrock test function using Algorithm~\ref{alg:1}. As the MCMC chain progresses and the diffusion model is repeatedly retrained, the acceptance rate increases well beyond 70\%. This indicates that the proposal diffusion model is closely approximating the target posterior, as explained in Sec.~\ref{sec:algorithm}.

\bibliographystyle{JHEP}
\bibliography{bibliography,stat}

\end{document}